\documentclass[conference]{IEEEtran}
\IEEEoverridecommandlockouts
\usepackage{cite}
\usepackage{amsmath,amssymb,amsfonts}
\usepackage{algorithm}
\usepackage{algorithmic}
\usepackage{graphicx}
\usepackage{textcomp}

\usepackage{colortbl} 
\usepackage[table]{xcolor} 
\usepackage{subcaption} 
\usepackage{url}

\def\BibTeX{{\rm B\kern-.05em{\sc i\kern-.025em b}\kern-.08em
    T\kern-.1667em\lower.7ex\hbox{E}\kern-.125emX}}
\begin{document}




\title{Bayesian Optimization for Fast Radio Mapping and Localization with an Autonomous Aerial Drone}

\author{
\IEEEauthorblockN{Paul S. Kudyba, Qin Lu, and Haijian Sun 
}
\IEEEauthorblockA{\textit{School of Electrical and Computer Engineering},
\textit{University of Georgia}, 
Athens, GA, USA \\
paul.kudyba@uga.edu,  qin.lu@uga.edu, hsun@uga.edu
}
}
\maketitle

\begin{abstract}
This paper explores how a flying drone can \mbox{autonomously} navigate while constructing a narrowband radio map for signal localization. As flying drones become more ubiquitous, their wireless signals will necessitate new wireless technologies and algorithms to provide robust radio infrastructure while preserving radio spectrum usage.
A potential solution for this spectrum-sharing localization challenge is to limit the bandwidth of any transmitter beacon.
However, location signaling with a narrow bandwidth necessitates improving a wireless aerial system's ability to filter a noisy signal, estimate the transmitter's location, and self-pilot to improve the location estimate.
By showing results through simulation, emulation, and a final drone flight experiment, this work provides an algorithm using a Gaussian process for radio signal estimation and Bayesian optimization for drone automatic guidance. This research supports advanced radio and aerial robotics applications in critical areas such as search-and-rescue, last-mile delivery, and large-scale platform digital twin development.
\end{abstract}


\section{Introduction}
Advances in aerial robotics offer engineers new opportunities to collect radio information with increased speed and cost efficiency \cite{spectrumcartography}. Within this revolution in capability lies unique engineering insights that could prove foundational to providing the next generation of wireless infrastructure. Such insights are already well established as areas of active wireless research, such as efficient and accurate radio cartography and radio mapping, which can be used to drive improved communication coverage and spectrum occupancy \cite{shresthaRadioMapEstimation2023}. Furthermore, radio data, once collected or during collection, can be used to drive further valuable inferences, such as transmitter location \cite{kwonRFSignalSource2023b}. However, these spatial radio mapping techniques rely on broadband transmissions, leaving the receiver to decompose the channel response, which compounds the effects of large- and small-scale fading \cite{spectrumcartography,matzFundamentalsTimeVaryingCommunication2011}. In contrast to a broadband transmitter, an extremely narrowband transmission (ratio of bandwidth to carrier frequency $\ll 1$) can be leveraged to simplify the channel modeling while also increasing the spectrum efficiency of a localization inference.
These combined radio mapping and localization inferences can be used in applications such as last-mile delivery, search-and-rescue, and development of digital twins \cite{kudybaThesis}.

A common element in many of these radio insights and inferences is the societal demand to simultaneously produce novel radio technologies and robust standards as infrastructure~\cite{AERPAW}. This nexus of aerial robotics, wireless technology, and machine learning allows researchers to explore new solution spaces and gain new insights that can, in turn, create new radio algorithm frameworks to be easily and rapidly deployed after thorough testing. Our experiment used the Aerial Experimentation and Research Platform for Advanced Wireless (AERPAW) to quickly iterate our algorithm design within an emulated digital twin environment and collect real-world data on the testbed. 

\begin{figure}[thbp]
\centerline{\includegraphics[width=1\linewidth]{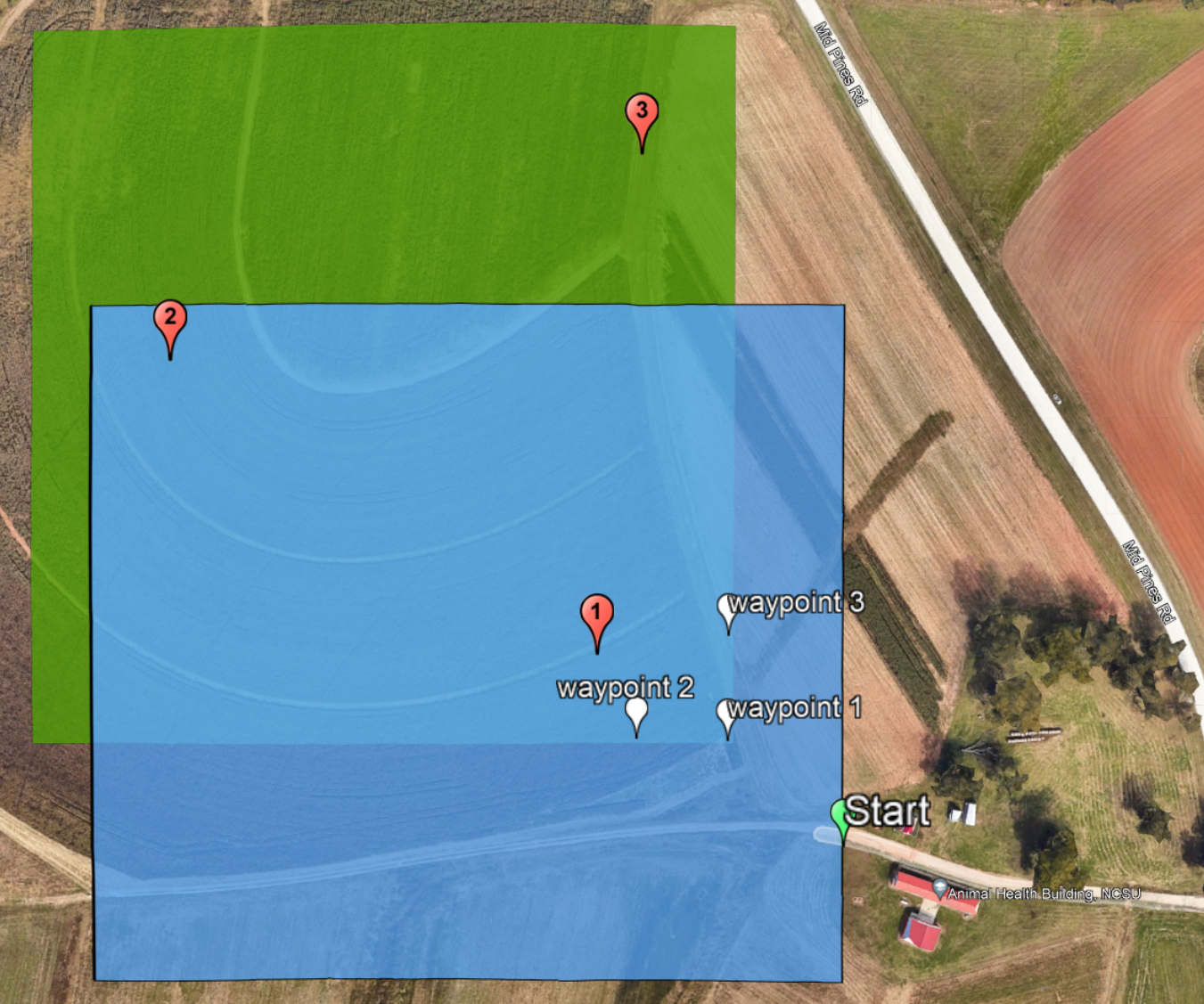}}
\caption{This figure shows the AERPAW boundaries for the AFAR challenge. The blue area indicates the aerial vehicle boundary limits. The green area indicates the possible locations of the rover. The numbered red points indicate where the rover was hidden, similar to Fig. 1 of \cite{kudybaUAVChallAERPAW}. The white waypoints indicate the locations sampled within our algorithm's startup routine for each run.}
\label{fig/overview}
\end{figure}

This paper brings forward a combined system implementation of aerial mobility, Software-Defined Radio (SDR), and machine learning into a fully deployable top-level algorithm that uses a Gaussian Process (GP) to produce a nonparametric 2D tomographic narrowband channel radio map on AERPAW.
As the aerial drone collects samples, the GP radio map forms a best approximation, or surrogate, of the unknown radio map spatial field.
A Bayesian Optimization (BO) Acquisition Function (AF) infers the best next location to collect a new radio sample.
Our methodology first simulated robot behavior with various pathloss functions, added signal noise, GP kernels, and BO AF.
A Laboratory testbed enabled signal analysis on the narrowband receiver and the development of a filter to reject deep fading. The final algorithm integrated all previous heuristics and analysis to run successfully on the AERPAW emulator and testbed.

The rest of the paper is structured as follows. The AERPAW challenge that enabled and guided this effort is described in Section \ref{sec:AFAR}. In Section \ref{sec:gp_for_radio_mapping}, a general structure of GP and how it is applied to construct an evolving radio map as samples are collected is presented. Section \ref{sec:kernel} gives a relation to how the kernel function and its optimization impact on the GP radio map. Section \ref{sec:bo} shows how once the GP has established a pathloss gradient, inference can be determined with BO. The methods used to create and deploy the final algorithm are established in Section \ref{sec:method} with the results shown in Section \ref{sec:results}. Finally, concluding remarks are given in Section \ref{sec:conclusion}. 

\subsection{AERPAW Find A Rover Challenge} \label{sec:AFAR} 
Participation in the first AERPAW challenge\footnote{\url{https://aerpaw.org/aerpaw-afar-challenge/}} provided an opportunity to address some of the active research questions mentioned above. The goal of the challenge was to use the platform emulator and testbed to locate a nonmoving ``rover" within three and ten minutes across an area of 19.7~acres using only a repeating BPSK pseudorandom pilot at 3.24~GHz with a 125~KHz bandwidth. A USRP B-210 mini SDR with a radio front end and filters was used to provide the best (lowest) mean error across three randomly chosen locations. Drone log estimates determined two results at the three- and ten-minute, marked from takeoff. Figure \ref{fig/overview} illustrates the boundaries and each of the final locations of the rover transmitter.
The final submissions were qualified by running them within the AERPAW emulator. If the emulator showed safe flight operation, the algorithm was used on the testbed to determine the final result.

To facilitate radio-based localization, the AERPAW team provided a channel-sounding script that used GNU Radio to interface with the SDR. This script ran a cross-correlation with the transmitted pseudorandom pilot and provided an output channel power in dB and a normalized SNR quality value.

\subsection{Gaussian Process for Spatial Coverage Radio Mapping} \label{sec:gp_for_radio_mapping} 
Kriging, emerging from practical use in geostatistics, was formalized into a canonical method called GP \cite{rasmussenGaussianProcessesMachine2006}. As is used here, a typical application employs a time-invariant surrogate spatial field mapping as stochastic processes or a Gaussian distribution of functions. The effectiveness of reducing a time dimension in a geostatistical context can be easily apparent by producing a cartography or spatial map. However, this particular dimension reduction assumption (e.g., a time-invariant channel) can cause issues in a mobility context. Knowingly, we use a more traditional movement-agnostic signal-processing receiver provided by AERPAW. Nevertheless, this exercise remains important in elucidating an assumption of static dimensionality in a novel system design context and its upstream implications for the system inference task. Any channel impairment processes not directly accounted for within the receiver design must be considered noise for upstream system insights. This includes temporal noise, such as multipath, Doppler, and sample jitter; even if some of these effects are relatively small, they are combined within our received signal. In this sense, the radio map we construct will forgo the time-varying dimension of a channel for a direct spatial alignment with radio communication as coverage and regard any temporal inconsistencies as noise.

\begin{figure}[tbp]
\centerline{\includegraphics[width=\linewidth]{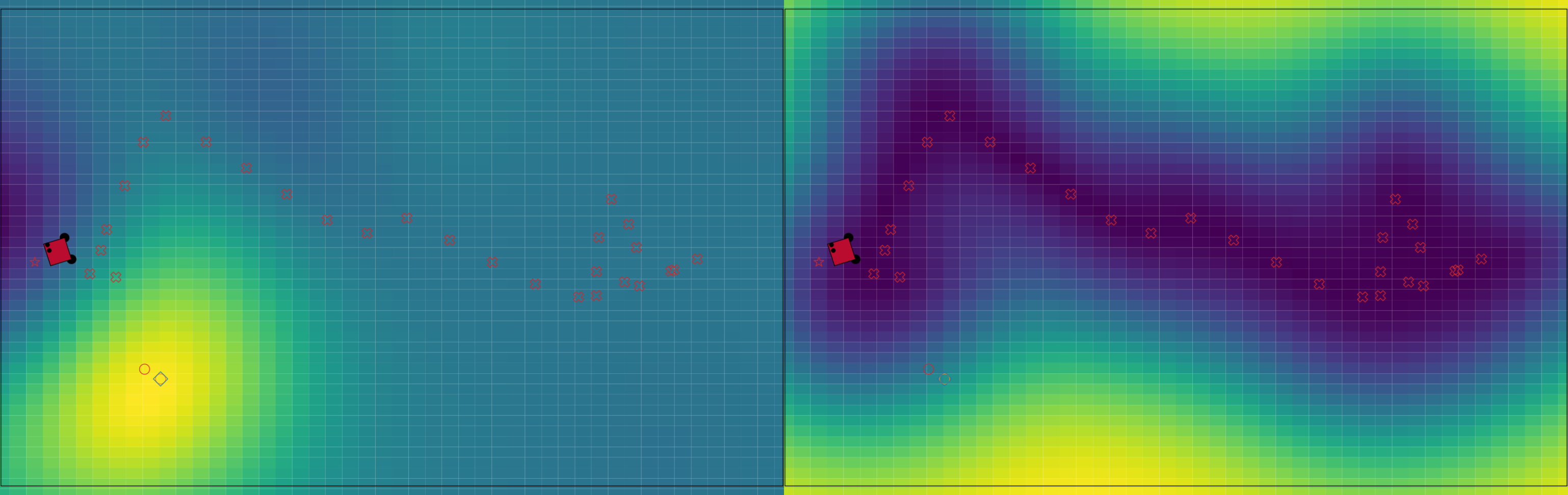}}
\caption{This figure shows the final results of a robotarium trial showing the GP mean (left) similar to Fig. 4a of \cite{kudybaUAVChallAERPAW}, and the GP uncertainty as a mean squared error (right). Yellow is a high value, and blue is a low value for both images.}
\label{fig/robotarium}
\end{figure}

With $\mathbf{x}_i$ being the coordinates of the $i$th location, $f(\mathbf{x}_i)$ is the {\it unknown} radio map density function, which, upon being corrupted by additive Gaussian white noise $v_i$, yields the noisy observation $p_i$. Mathematically, this can be described by the following observation model
\begin{equation}
p_i = f(\mathbf{x}_i) + v_i\;.
\label{eq:gp_noise} 
\end{equation}
To efficiently learn the radio map density function $f$, we will model it with a GP prior, namely
\begin{equation}
f(\cdot) \sim \mathcal{GP}(m(\cdot), k(\cdot, \cdot))\;
\label{eq:gp} 
\end{equation}
where the mean function $m(\cdot)$ is set to a constant $g$, representing the default channel gain corresponding to no signal or receiving incoherent noise, and the covariance or kernel function $k(\cdot,\cdot)$ measures the correlation or pairwise similarity of any two function values through their inputs, i.e., ${\rm cov}[f({\bf x}), f({\bf x}')]:=k(\mathbf{x}, \mathbf{x}')$. Here, the covariance function $k$ is characterized by some hyperparameters, which can be estimated via the marginal likelihood maximization as described in the next subsection.

Notably, this GP prior describes a probability distribution of the radio map functions $f(\cdot)$, which, upon collecting the set of data sampling $\{({\bf x}_i,p_i)_i\}$, yields the function posterior
\begin{equation}
   p(f({\bf x})|\{({\bf x}_i,p_i)_i\}) = \mathcal{N}(f({\bf x}); \mu({\bf x}), \sigma({\bf x})) \label{eq:post}
\end{equation}
with the mean $\mu({\bf x})$ and standard deviation $\sigma({\bf x})$ of the estimated radio map at any location ${\bf x}$.
%
In our case, $f(\cdot)$ resembles a pathloss, and our expectation is that it should match an exponential decay, but the actual decay exponent remains an unknown parameter (due to the novel system and environment). Thus, this pathloss exponent will be found at run-time as a hyperparameter.

With the correct kernel and hyperparameters, the true spatial field channel, $f(\cdot)$, will be drawn from the GP distribution of radio maps as the mean. This regression, with an output indicating the most probable radio map and associated uncertainty, is how we generate a parsimonious and robust atemporal channel estimate map. Additionally and intuitively, the location of the signal peak within the estimated radio map, when all the dimensionality reduction assumptions and noise rejections hold, is the transmitter's true location.
\subsection{Kernel Functions and Maximum Likelihood Algorithm} \label{sec:kernel}
The positive and semi-definite kernel function $k(\cdot,\cdot)$ plays a performance-critical role in GP modeling.  Notably, the commonly used kernel is the Rational Basis Function (RBF) function, whose expression is given by
\begin{equation}
k_{\mathrm{RBF}}({\bf x},{\bf x}^{\prime})=\gamma^{2}\exp\biggl(-\frac{\|{\bf x}-{\bf x}^{\prime}\|_2^{2}}{2\ell^{2}}\biggr)\;.
\label{eq:gp_kernel_se} 
\end{equation}
Intuitively, the value of $k({\bf x}, {\bf x}')$ will become larger as the distance between ${\bf x}$ and ${\bf x}'$ becomes smaller.
This inherently produces a distribution of infinitely smooth functions that exponentially decay as expected from any free space pathloss function. However, the RBF, used alone, would also fit any noise incorrectly to the posterior as part of the pathloss. Fortunately, with the addition of a white noise kernel, each posterior sample will now include a hyperparameter of noise within each sample collected, allowing the entire radio map distribution to have more flexibility for inconsistencies in the input sampling. Combined, the starting kernel Python code is given as \texttt{kernel = (C(1) * RBF(0.00276, (0.001,0.004)) + WhiteKernel(noise\_level=0.33)}. This shows an RBF with a $\ell$ value of 0.00276 being bounded between 0.001 and 0.004.

Each kernel comes with a specific set of hyperparameters; see, e.g., the scaling factor $\gamma$ and lengthscale $\ell$ in the RBF kernel in Eq. \ref{eq:gp_kernel_se}. These hyperparameters are obtained by maximizing the marginal likelihood via the Limited-memory Broyden-Fletcher-Goldfarb-Shanno with Bound constraints \mbox{(L-BFGS-B)} algorithm. Here, the bounding is used to ensure an applicable resulting lengthscale at all times during the flight. 

\subsection{Navigation with Acquisition} \label{sec:bo}
The GP modeling yields a posterior distribution of the radio map function as in~\eqref{eq:post}. A consideration for mapping/localization is the time penalty, opportunity cost, or regret to produce each reliable sample~\cite{santosMultirobotLearningCoverage2021}; the drone simply cannot sample the entire field, and yet it needs to know the best next location to fly towards and sample. The GP is also very likely to change as samples are added, especially at the beginning of any flight. These properties disallow many traditional forms of optimization; however, because our radio map has an associated uncertainty provided by our GP kernel function, BO can be used to infer an optimal future sample location. This future location will be selected from the GP ``black box" surrogacy, from which a maximum can be chosen according to an AF. BO, therefore, links our GP output of a mean ($\mu$) and uncertainty ($\sigma$) to a balanced tradeoff of where to collect a new sample.
BO accomplishes this with an AF that balances two objectives: \textit{exploring} uncertain positions to find the global optimum and \textit{exploiting} the best yet known mean value. 
Here, we adopt the upper confidence bound (UCB) based AF, given by 
\begin{equation}
\mathrm{UCB}({\bf x}; \beta_n) = \mu({\bf x}) + \beta_n \sigma({\bf x})
\label{eq:ucb} 
\end{equation}
where $\beta_n$ is a non-negative exponentially decreasing sequence updated after collecting each sample. We choose $\beta_n = d^s$ with $0<d<1$. This choice, in effect, reduces the exploration term as the flight progresses and samples are taken.

\section{Experimental Methodology} \label{sec:method} 
The production of a final AFAR submission consisted of three stages of research and development: simulation via Robotarium~\cite{pickemRobotariumRemotelyAccessible2017}, receiver testing with an in-lab setup, and emulation with AERPAW. The results of each stage gave valuable insight into the next, building into a cohesive and successful deployment.


\begin{figure*}[htb]
\centerline{\includegraphics[width=\linewidth]{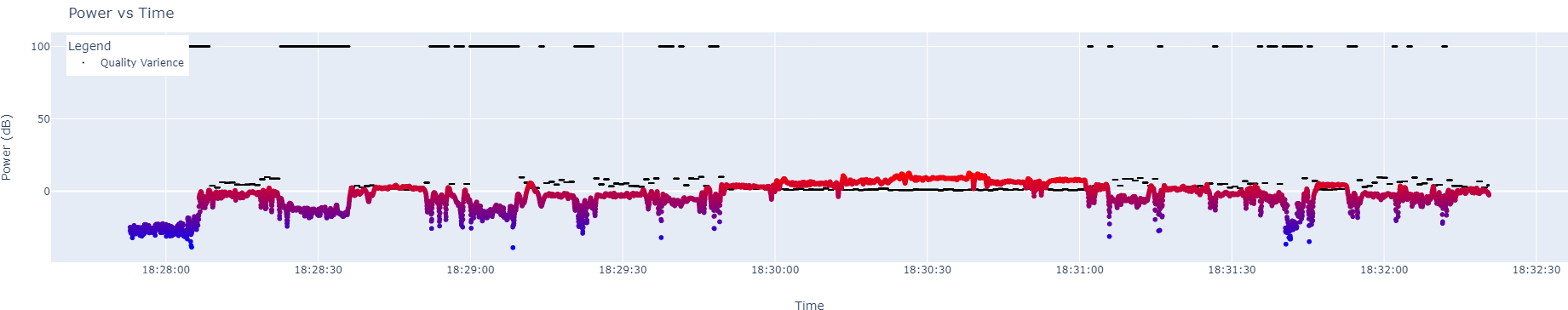}}
\caption{This figure shows the output of the quality-variance filter along with the signal intensity given from the AERPAW narrowband channel sounder script. This data was collected in a lab setting that resembles the AERPAW setup without any RF front end. The receiver was placed on a moving platform, which was moved from 16~ft to 4~ft and then back to 16~ft from the transmitter.
This figure is adapted from Fig. 4.2f of \cite{kudybaThesis}.
}
\label{fig/filter}
\end{figure*}

A 2D robotic simulator was used to analyze the expected self-guided robot behavior in controlled scenarios, and pathloss functionality was added to facilitate autonomous development. For example, a 2D normal distribution as a basic pathloss function allowed control of the mean as a transmitter's location and the standard deviation to control the decay. The platform enabled many tests with varying pathloss models and signal-to-noise levels. It was also seen that two predefined routines before and after autonomous BO were beneficial to stability, efficiency, and accuracy. The robot would start with a predefined circle routine wide enough to establish a stable gradient and resolve any noise for BO to give stable guidance. Secondly, after a stable global maximum was repeatedly chosen as the optimal location, a routine could be triggered to circle that point, further refining the transmitter's location. Lastly, the simulation allowed experimentation with different AF, leading to the selection of UCB for the final deployment. Figure \ref{fig/robotarium} shows the mean and uncertainty radio maps resulting from BO with uniform noise added to the robot's receiver. A 2D grid was constructed to facilitate getting discrete spatial field points of the GP mean and maximum squared error. The AF could be computed from this grid, and the maximum would then be selected as the new target waypoint. 

To ensure that the GP receives no outlier radio data that could not be described as Gaussian noise, a non-linear filter was shown to be necessary from laboratory testing and data from a previous testbed run. A large discriminator was observed by binning the receiver quality signal and taking the variance of these readings. By removing any readings with an excessive quality-variance, large signal noise swings could be rejected as invalid samples and excluded from the GP. However, this threshold needed to be determined at flight time due to limited empirical data. Laboratory testing also indicated that receiver movement influenced the signal's tendency to encounter noise. The binning of the quality metric in a lab setup where the receiver periodically moves towards and away from the transmitter is shown in Figure \ref{fig/filter}.

A final AERPAW algorithm was then created to perform emulation testing and for the final deployment on AERPAW. Two grids are created using the latitude and longitude for both the flying boundary and the rover location boundary. The flying boundary grid is then used for all navigational inputs, and the rover's final estimates are given from the second rover grid. Both grids are sampled from the same GP.
The drone then takes off to 40~m for the entire trial. Once at the correct altitude, the drone attempts to take the first radio sample binning data for 6~s while stationary and always facing northwest. If the quality-variance of the first sample is higher than the threshold, the sample is retaken with an exponentially higher set threshold until this first sample is accepted. The drone then flies to three predefined waypoints shown in Figure \ref{fig/overview}. These waypoint samples give a stable GP and BO for autonomous flight. The drone then flies its mission according to the same AF UCB policy chosen in the simulation. Two possible criteria can trigger an auxiliary routine that creates a circle of waypoints from the drone's current position (observing the boundary limits). The first criterion is that the AF has repeatedly chosen very spatially similar points to investigate. This is similar to the simulation ending routine, which indicates the drone is close to the maximum channel gain. The second is if the drone rejects a maximum number of measurements throughout the mission. This is a failsafe that was seen to benefit the mission while testing within the AERPAW emulator. If the drone cannot navigate autonomously to points that provide valid radio samples for the GP, a circular routine of waypoints ensures that the radio will be given a broad spatial set of sample locations, which might recover a gradient towards the transmitter.
During the mission, a mission timer tracks the duration and logs the final predicted location from the rover grid space for the three- and ten-minute estimates.

\section{Results and Analysis} \label{sec:results}

{\tiny 
\begin{table*}[tbp]
\centering
\caption[Experimental simulation and testbed results]{The three-run simulation and testbed results. The bracketed number indicates the trial number corresponding to the hidden locations given in Figure \ref{fig/overview}.}
\begin{tabular}{|c|c|c|c|c|c|c|c|c|}
\hline
\rowcolor{gray!50}
3-min (T.1)  & 3-min (T.2)  & 3-min (T.3)  & 3-min Average & 10-min (T.1) & 10-min (T.2) & 10-min (T.3) & 10-min Average \\ \hline
\rowcolor{gray!25}
\multicolumn{8}{|c|}{Simulation Results} \\ \hline
71.7m  & 265m   & 174m   & 170.23m      & 52.8m  & 17m    & 19m    & 29.6m         \\ \hline

\rowcolor{gray!25}
\multicolumn{8}{|c|}{Testbed Results} \\ \hline
40.5m  & 239.1m & 232.4m & 170.7m       & 41.6m  & 27.8m  & 130.4m & 66.6m         \\ \hline

\end{tabular}

\label{tab:results}
\end{table*}
} 

Table \ref{tab:results} presents the results from the AERPAW emulator and testbed environments. The first testbed mission shows how quickly the GP can fit a transmitter that is relatively close (125~m) from the drone's starting location. However, the drone could not resolve a better 10-minute location estimate due to receiver gain saturation. While a consistently high gain is generally desired for communication, corresponding to an ideal channel, in this case, the signal 'clipping' makes a spatial 'plateau.' This spatial feature makes resolving a transmitter location nearly impossible when within this region. Seeing this within the emulator, the radius for the auxiliary circle was set expecting to encounter this signal behavior; however, in this case,  the auxiliary path was not wide enough to provide sufficient localization information. Additionally,  critical points were missed on the northwest side of the auxiliary path due to noise and correctly excluded by the quality-variance filter.

\begin{figure}[!htb]
    \centering
    \begin{subfigure}[b]{\linewidth}
        \centering
        \includegraphics[width=0.8\textwidth]{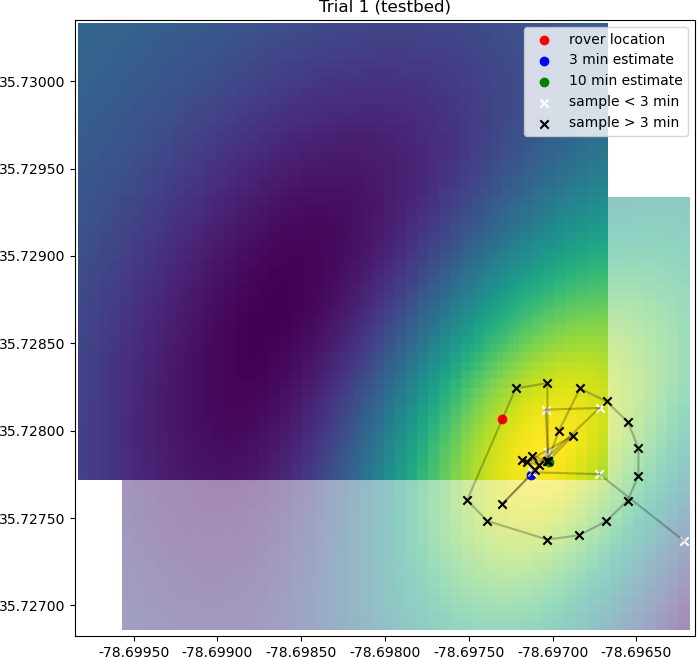}
        \caption{}
        \label{fig/run2_gp_1}
    \end{subfigure}
    \begin{subfigure}[b]{\linewidth}
        \centering
        \includegraphics[width=0.8\textwidth]{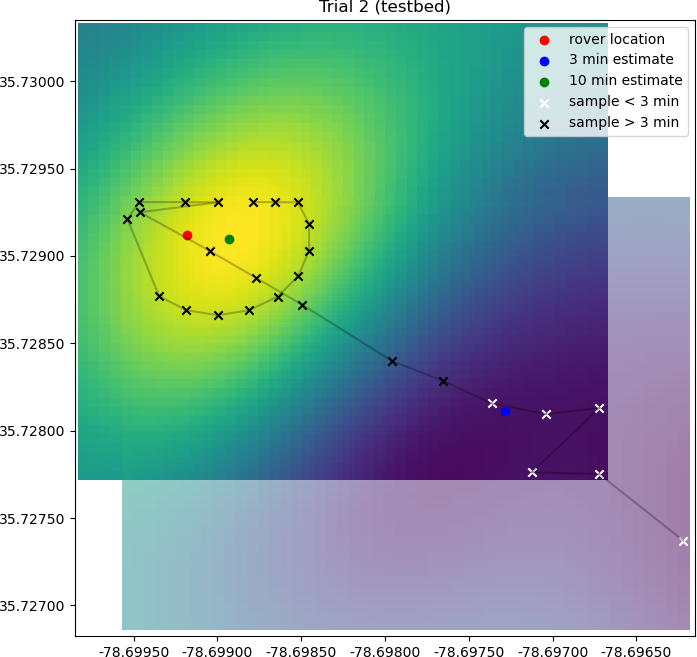}
        \caption{}
        \label{fig/run2_gp_2}
    \end{subfigure}
    \caption{This figure shows the first and second run reconstructions of the final GP estimates over the search spaces shown in figure \ref{fig/overview}. The actual rover location is shown as a red dot. The green dot shows the final ten-minute estimate, and the blue dot indicates the three-minute estimate. This GP reconstruction uses the final kernel hyperparameters indicated by the vehicle log.
    Part of this figure is adapted from Fig. 4c of \cite{kudybaUAVChallAERPAW}.
    }
    \label{fig/run2_gp_combined}
\end{figure}

The second trial, shown in Figure \ref{fig/run2_gp_2}, gives the best 10-minute result of 27.8~m. The 3-minute result did not provide quite enough time for the drone to establish a proper gradient and travel to the transmitter 330~m from takeoff. However, the drone was able to establish a proper optimal trajectory toward the transmitter and perform the auxiliary path maneuver while staying within the boundary conditions. The auxiliary path was wide enough to provide additional localization to the GP, and the quality-variance filter was working appropriately to filter excessive noise. This result shows the ideal behavior of the drone in this situation. To provide a better three-minute GP estimate with this starting routine, a kernel that supplied more information about the signal from this starting distance (as a prior) would need to be considered.

During the third trial, the drone encountered a confluence of issues, which resulted in degraded estimation performance. However, the mission was not a failure; the drone was able to produce locations for both timed estimates from a starting distance of 286~m. Log analysis revealed that the quality-variance filter threshold (set with the first sample) was set to exclude any variance above 35. This was different from the second run, which allowed samples with a less strict value of 67. This disallowed many of the first critical direction-finding waypoints from being included within the GP. The drone then generated the auxiliary path due to missing too many sample measurements. From this waypoint path, two samples were accepted. These samples greatly increased the accuracy of the location estimate. The drone reestablished an optimal path but was unable to collect any more samples due to the time limit.

\section{Conclusion} \label{sec:conclusion} 
The results of the three trials show significant robustness in the GP-based BO approach for drone-based localization. The quality-variance filter identified and removed outlier noise values from the narrowband receiver. Unfortunately, the drone was unable to find the correct filter parameter value from the start of the third trial. This alone shows a necessity for bridging the gap between the AERPAW emulator and real-world radio data. If the emulation noise and the testbed had a reliable digital-twin agreement, the filter parameter could be safely studied and set within the emulator, similar to how the auxiliary path area was tested and adopted. This would save mission time and prevent mission-hazardous calibrations.
Additionally, by more closely integrating the digital twin emulation/testbed environment, various ``black box" pathloss regression estimators for mapping and localization could be trained and trialed within an emulated environment with accurate synthetic radio data.
%

\vspace{12pt}

\end{document}